\documentclass[prl,twocolumn,floatfix,preprintnumbers]{revtex4}
\usepackage{graphicx}
\usepackage{dcolumn}
\usepackage{bm}
\usepackage{color}

\providecommand{\adsurl}[1]{\href{#1}{ADS}}

\def\lsim{\mathrel{\mathop
  {\hbox{\lower0.5ex\hbox{$\sim$}\kern-0.8em\lower-0.7ex\hbox{$<$}}}}}
\def\gsim{\mathrel{\mathop
  {\hbox{\lower0.5ex\hbox{$\sim$}\kern-0.8em\lower-0.7ex\hbox{$>$}}}}}

\begin{document}
\newcommand{\mincir}{\raise
-2.truept\hbox{\rlap{\hbox{$\sim$}}\raise5.truept 
\hbox{$<$}\ }}
\newcommand{\magcir}{\raise
-2.truept\hbox{\rlap{\hbox{$\sim$}}\raise5.truept
\hbox{$>$}\ }}
\newcommand{\minmag}{\raise-2.truept\hbox{\rlap{\hbox{$<$}}\raise
6.truept\hbox
{$>$}\ }}

\newcommand{\half}{{1\over2}}
\newcommand{\bk}{{\bf k}}
\newcommand{\Ocdm}{\Omega_{\rm cdm}}
\newcommand{\ocdm}{\omega_{\rm cdm}}
\newcommand{\OM}{\Omega_{\rm M}}
\newcommand{\OB}{\Omega_{\rm B}}
\newcommand{\oB}{\omega_{\rm B}}
\newcommand{\OX}{\Omega_{\rm X}}
\newcommand{\cltt}{C_l^{\rm TT}}
\newcommand{\clte}{C_l^{\rm TE}}
\newcommand{\mwdm}{m_{\rm WDM}}
\newcommand{\mnu}{\sum m_{\rm \nu}}
\newcommand{\etal}{{\it et al.~}}
 \newcommand{\lya}{{Lyman-$\alpha$~}}
\newcommand{\gad} {{\small {GADGET-2}}\,}
\input epsf

\title{How cold is cold dark matter? Small scales constraints from the
flux power spectrum of the high-redshift \lya forest} \author{Matteo
Viel,$^{1,2}$
George D.~Becker, $^3$
James S.~Bolton, $^4$
Martin G.~Haehnelt,$^5$ 
Michael Rauch, $^6$
Wallace L.W. Sargent $^3$}

\affiliation{
$^1$ INAF - Osservatorio Astronomico di Trieste, Via G.B. Tiepolo 11,
I-34131 Trieste, Italy \\
$^2$ INFN/National Institute for Nuclear Physics, Via Valerio 2,  
I-34127 Trieste, Italy \\
$^3$ Palomar Observatory, California Institute of Technology, Pasadena, CA 91125, USA\\
$^4$ Max-Planck-Institut f\"ur Astrophysik, Karl-Schwarzschild-Strasse 1, D-85741, Garching, Germany \\
$^5$ Institute of Astronomy, Madingley Road, Cambridge CB3 0HA, United Kingdom\\
$^6$ Carnegie Observatories, 813 Santa Barbara Street, Pasadena, CA 91101, USA
}
\date{\today}

\begin{abstract}
We present  constraints on the mass of warm dark matter (WDM)
particles derived from the \lya flux power spectrum of 55 high-
resolution HIRES  spectra at $2.0 < z < 6.4$.  From the
HIRES spectra, we obtain a lower limit of $\mwdm \gsim 1.2$ ~keV
($2\sigma$) if the WDM consists of early decoupled thermal relics and
$\mwdm \gsim 5.6$~keV ($2\sigma$) for sterile neutrinos. Adding the
Sloan Digital Sky Survey \lya flux power spectrum, we
get $\mwdm \gsim 4$ ~keV and $\mwdm \gsim 28$ ~keV (2$\sigma$) for
thermal relics and sterile neutrinos. These results improve previous
constraints by a factor two.
\end{abstract}


\maketitle
 
{\bf {Introduction.}}  Warm dark matter (WDM) has been advocated in
order to solve the apparent problems of standard cold dark matter
(CDM) scenarios at small scales (e.g. \cite{bode}), most notably: the
excess of the number of galactic satellites, the cuspiness and high
(phase space) density of galactic cores and the large number of
galaxies filling voids.  These and other problems could be alleviated
if the dark matter (DM) is made of warm instead of cold particles. The
main effect of the larger thermal velocities would be to suppress
structures below the Mpc scale.  However, poorly understood
astrophysical processes governed by the baryonic component of
galaxies, along with numerical, observational and theoretical aspects
\cite{Wang:2007he,strigari2,fornax}, have also to be considered in
order to reliably model the spatial distribution of dark matter at
small scales. The \lya absorption produced by the intervening neutral
hydrogen in the spectra of distant quasars (QSOs), the so called \lya
forest, is a powerful tool for constraining dark matter properties.
It probes the matter power spectrum in the mildly non-linear regime
over a large range of redshifts ($z=2-6$) down to small scales
($1-80\, h^{-1}$ Mpc) \cite{Seljak:2006bg}.  In previous work,
Ref. \cite{Viel:2005qj} used two samples of high-resolution QSO \lya
forest spectra at $z\sim2.5$ to set a lower limit of 550 eV for the
mass of a thermal warm dark matter candidate (2 keV in case of a
sterile neutrino).  More recently, Ref. \cite{Seljak:2006qw} and
Ref. \cite{Viel:2006kd}, using the Sloan Digital Sky Survey (SDSS) QSO
data set at higher redshifts and different methods significantly
improved this limit by a factor $\sim 4$.  Among the possible WDM
candidates, the most promising appears to be a sterile neutrino with a
mass in the keV range, which could be part of many particle physics
models with grand unification (e.g. \citep{pulsar}). Because of a
non-zero mixing angle between active and sterile flavor states, X-ray
flux observations can also constrain the abundance and decay rate of
such DM particles (e.g. \cite{xrayall}). The constraints from \lya
forest data and those from the X-ray fluxes of astrophysical objects
together put considerable tension on the parameter space still allowed
for a sterile neutrino particle with the phase-space distribution
proposed by Dodelson \& Widrow (DW) \cite{dw}, although other
non-standard scenarios must be explored \cite{kusenko}. Here, we will
use a new large set of high-resolution \lya forest spectra in order to
improve limits on the mass of a WDM particle.

{\bf {Data sets.}} We use two different data sets: $i)$ the high
resolution HIRES data set presented in \cite{Becker} which consists of
55 QSOs spanning the range $2.0 < z < 6.4$; $ii)$ the SDSS \lya forest
data of McDonald et al. \cite{McDonald:2004eu}, which consists of
$3035$ quasar spectra at low resolution ($R\sim 2000$) and low
signal-to-noise spanning a wide range of redshifts ($z=2.2-4.2$). We
have calculated the flux power spectrum from the HIRES data for 4
redshift bins with median redshifts $z=2.5,3.5,4.5,5.5$ and 20
logarithmically spaced bins in wavenumber spanning $0.002 <k$ (s/km)$
<0.287$, Note, however, that we use only 12 bins in the range $0.003
<k$ (s/km)$ <0.077$ for our present analysis (a total of 48 points).
The total redshift path of the HIRES sample is $\Delta z=29.1$.  The
redshift path of the lowest redshift bin is comparable to that of the
LUQAS sample of high-resolution spectra ($\Delta z=13$) which we used
previously \citep{kimcroft}, while the other three redshift bins have
smaller redshift paths of approximately equal length.  Note that our
previous work based on the LUQAS sample used only the range $0.003 <k$
(s/km)$ <0.03$.  For the SDSS data set we use 132 flux power spectrum
measurements $P_F({\bf{k}},z)$ that span 11 redshift bins and 12
$k-$wavenumbers in the range $0.00141< k$ (s/km)$ <0.01778$ (roughly
corresponding to scales of 5-50 comoving Mpc). Since the HIRES spectra
have higher resolution than the SDSS spectra we can use the flux power
spectrum obtained from the HIRES data to extend our analysis to
smaller scales. We have removed the damped \lya systems (DLAs) from
the observed HIRES spectra but decided not to attempt to remove the
metal lines. We rely on the measurement of the contribution of
metal absorption to the flux power spectrum at $z=2.13$ by
\cite{kimcroft}.  We use this estimate to correct our measurement of
the flux power spectrum at $z=2.5$, and apply a smaller correction for
the measurement at $z=3.5$. For the two highest redshift bins we do
not apply any correction to our measurements of the flux power
spectrum as the metal contribution should be very small at these
redshifts. The contribution of the metal absorption to the flux power
spectrum at the scales considered here depends only weakly on
wavenumber and is degenerate with the value of the assumed effective
optical depth \cite{Viel:2005ha}. Marginalizing over the observed
range of the effective optical depth in our parameter analysis will
(implicitly) account for a possibly different contribution to the flux
power spectrum than we have assumed. We have tested the effect of
continuum fitting errors by calculating the flux power spectrum of
each observed \lya forest spectrum at $z>4$ for 100 continuum fit
realizations where we have adjusted the continuum level of the
observed spectra randomly by a factor 1+$\epsilon$ with $\epsilon$ in
the range $[-0.1,0.1]$. This should be a reasonable estimate of the
effect of continuum fitting errors at high redshift.  At smaller
redshifts the continuum fitting errors should be below 4\%. The
variations of the flux power for these 100 realizations lies well
within the statistical errors of the flux power spectrum of our full
sample and estimates differ by less than 5\% from the case with
$\epsilon=0$. We have decided not to try to explicitly account for
continuum fitting errors. The covariance matrix of the flux power
spectrum for the HIRES data set was calculated with a jack-knife
estimator.

{\bf {Method.}} Modeling the flux power spectrum of the \lya forest
for given cosmological parameters down to the required small scales is
not straightforward and accurate numerical supercomputer simulations
are required.  Here, we model the flux power spectrum with full
hydro-dynamical simulations using a second order Taylor expansion
around a best fitting model.  This allows us to obtain a reasonably
accurate prediction of the flux power spectrum for a large range of
parameters, based on a moderate number of simulations
\cite{Springel:2005mi}. The method has been first introduced in Ref.\
\cite{Viel:2005ha} and we refer to this work for further details. Note
that while in Ref.\ \cite{Viel:2005ha} the prediction for the flux
power were made using a first order Taylor expansion, here the
expansion is made to second order: i.e. the parameter dependence of
the flux power spectrum $P_F({\bf k},z,{\bf p})$ is locally described
by a 2nd order polynomial function for any redshift $z$, set of
wavenumbers ${\bf k}$ and cosmological or astrophysical parameters
${\bf p}$.  In this work, we use as input in the hydro-dynamical
simulations a linear matter power spectrum as in
\cite{CDW,Viel:2005qj} and we assume that the sterile neutrino
phase-space distribution is equal to that of active neutrinos
multiplied by a suppression factor. Deviations from this first-order
approximation were computed in \cite{Abazajian:2005gj}, but typically
these corrections lower the bounds on the warm dark matter mass by
only 10\% \cite{Seljak:2006qw}.

For our best estimate of the flux power spectrum of our fiducial model
we used a simulation of a box of length 60 $h^{-1}$ comoving Mpc with
$2\times400^3$ gas and cold DM particles (gravitational softening 2.5
$h^{-1}$ kpc). The fiducial flux power spectrum has been corrected for
box size and resolution effects. Note that resolution corrections are
large at high redshift and redshift/scale dependent: they reach 50\%
(300\%) at the smallest scales at $z=4.5$ ($z=5.5$).  We performed a
number of additional hydrodynamical simulations with a box size of 20
$h^{-1}$ comoving Mpc and $2\times 256^3$ gas and DM particles
(grav. soft. 1 $h^{-1}$ kpc) for WDM models with a (thermal) warm dark
matter of mass $m_{\rm WDM}=1,4,8$ keV, to calculate the derivatives
of the flux power spectrum with respect to changes of the WDM particle
mass and other astrophysical and cosmological parameters of interest
in this analysis.  We have checked the convergence of the flux power
spectrum on the relevant scales using several additional simulations
(following the approach of \cite{McDonald03}) with $2\times 256^3$ gas
and DM particles and box sizes of 10 $h^{-1}$ Mpc (grav. soft. 2
$h^{-1}$ kpc), 5 $h^{-1}$ Mpc (grav. soft. 1 $h^{-1}$ kpc) and a 5
$h^{-1}$ Mpc simulation with $2\times 448^3$ (grav. soft. 0.27
$h^{-1}$ kpc) .

\begin{figure}[h!]
\begin{center}
\includegraphics[angle=0,width=8.cm,height=8.cm]{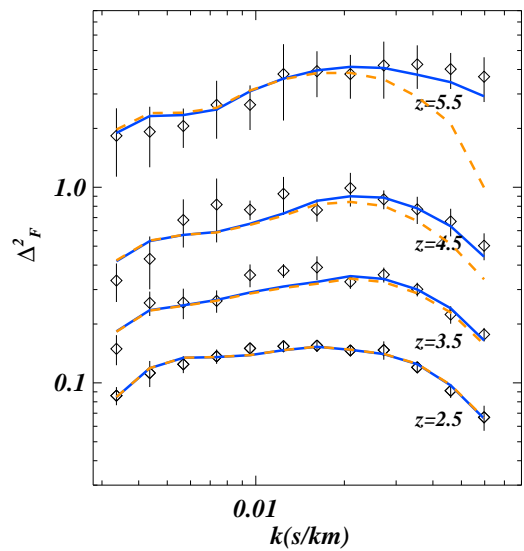}
\end{center}
\vspace{-0.5cm}
\caption{\label{f1} Flux power spectrum of the HIRES data set at
different redshifts and best fit models (solid curve) with $m_{\rm
WDM} = 8$ keV  and a model with $m_{\rm WDM} = 2.5$ keV
(dashed curve).}
\end {figure}

We then used a modified version of the code CosmoMC
\cite{Lewis:2002ah} to derive parameter likelihoods from the the HIRES
and SDSS \lya data. For the HIRES data, we had a set of 18
parameters: 7 cosmological parameters; 6 parameters describing the
thermal state of the Intergalactic Medium: parameterization of the gas
temperature-gas density relation $T=T_0(z)(1+\delta)^{\gamma(z)-1}$ as
a broken power law at $z=3$ with the two astrophysical parameters
$T_0^A(z)$ and $\gamma^A(z)$ describing the amplitude; 4 parameters
describing the evolution of the effective optical depth with redshift
(slope $\tau^S_{\rm eff}$ and amplitude $\tau^A_{\rm eff}$ at $z=3$
and $z=5$) since a single power-law has been shown to be a poor
approximation over this wide redshift range (see \cite{Becker}); one
parameter describing the spatial fluctuations of the 
Ultra-Violet (UV) background $f_{\rm UV}$.

In estimating the effect of UV fluctuations on the flux power we adopt
a conservative approach and consider the model of \cite{bolton}, where
the UV fluctuation have a large impact on the flux.  The model assumes
that the UV background and its spatial fluctuations are produced by
Lyman-Break galaxies and QSOs and uses as input the QSOs and
Lyman-Break luminosity functions at $z=3.5,4,5,6$. At $z=5.5$ the flux
power in the model with UV fluctuations is larger by 4\% at the
largest scales increasing to 20\% at $k=0.2$ s/km (not included in the
analysis), compared to the case without UV fluctuations.  At $z=4$ and
$z=3.5$ the only differences arise at scales $k>0.3$ s/km, which are
not considered in the present analysis. Further details on the UV
model can be found in \cite{bolton} (but see other approaches
\cite{UV}). We parameterize the effect of UV fluctuations on the flux
power with a multiplicative factor $f_{UV}$ constrained to be in the
range $[0,1]$. For the SDSS data we have used a total of 28
parameters: 15 parameters used for the HIRES spectra (without $f_{ \rm
UV}$ and the two parameters describing the effective optical depth
evolution at $z=5$) plus 13 noise-related parameters: 1 parameter
which accounts for the contribution of DLAs and 12 parameters
modelling the resolution and the noise properties of the SDSS data set
(see \cite{McDonald:2004xn}).  We do not address the role of different
reionization scenarios on the flux power. To do this
self-consistently would require radiative transfer simulations beyond
present numerical capabilities and the effect
of the reionization history should be subdominant and degenerate with
the thermal state of the gas. In computing the likelihood a crucial
input is the covariance matrix of the two data sets. The covariance
matrix of the SDSS flux power is provided by the authors of
\cite{McDonald:2004eu}.  We found the covariance matrix of our HIRES
data set to be rather noisy (especially at high redshift), preventing
a reliable inversion. To overcome this problem we use the suggestion
of \cite{Lidz:2005gz}.  We regularize the observed covariance matrix
using the correlation coefficients as estimated from the simulated
spectra, $cov_d(i,j)=r_s(i,j)\sqrt{cov_d(i,i)cov_d(j,j)}$ with
$r_s(i,j)=cov_s(i,j)/\sqrt{cov_s(i,i)cov_s(j,j)}$, where $cov_s$ and
$cov_d$ are the covariance matrices of the observed and simulated
spectra, respectively. Note that this procedure implicitly assumes
that observed and simulated data have similar covariance properties.
We have  applied moderate priors to the thermal history to
mimic the observed thermal evolution as in \cite{Viel:2005ha} and a
prior on the Hubble constant (72 $\pm 8$ km/s/Mpc), but note that the
final results for the mass constraint are not
affected by these priors.

{\bf {Results.}}  In Figure \ref{f1} we show the best fit model for
the HIRES data set (continuous curve, $m_{\rm WDM}=8$ keV) and a model
with a smaller mass for the thermal WDM particle (dashed line, $m_{\rm
WDM}=2.5$ keV ). The constraining power of the small scales at high
redshift is immediately evident.  The $\chi^2$ value of the best fit
model is $\sim 40$ for 36 d.o.f.  and with a probability of 16\% this
is a reasonable fit. As noted in Ref. \cite{Seljak:2006qw} at high
redshifts, the mean flux level is lower and the flux power spectrum is
closer to the linear prediction making the flux power data points very
sensitive to the free-streaming effect of WDM. We confirm that there
are no strong degeneracies between $m_{\rm WDM}$ and the other
parameters, demonstrating that the effect of a WDM particle on the
\lya flux power is unique, and that the other cosmological and
astrophysical parameters considered here cannot mimic its effect.

The $2\sigma$ lower limits for the mass of the warm dark matter
particle are: 1.2 keV, 2.3 keV and 4 keV, for the HIRES, SDSS and
SDSS+HIRES data sets, respectively.  The corresponding limits for DW
sterile neutrino are: 5.6, 13, and 28 keV (see \cite{Viel:2005qj} for
how the masses are related for the two cases).  The $\chi^2$ of the
best fit model of the joint analysis $\sim 198$ for 170 d.o.f. which
should occur in 7\% of the cases. The sample of HIRES spectra improves
our previous constraint from high-resolution spectra obtained from the
LUQAS sample by a factor two.  Dropping the highest redshift bin
($z=5.5$) weakens the limit to 0.8 keV (3.3 keV) for the mass of a
thermal (sterile) neutrino.  The SDSS data alone is still more
constraining than the HIRES data alone, due to the smaller statistical
errors of the SDSS flux power spectrum and the finer coverage of a
large redshift range which helps to break some of the degeneracies
between astrophysical and cosmological parameters.  Combining the SDSS
data and the HIRES results in an overall improvement of a factor $\sim
2$ and gives the strongest limits on the mass of WDM particles from
\lya forest data to date.  In Table 1 we summarize the constraints
obtained for the most relevant astrophysical and cosmological
parameters (1$\sigma$) for our analysis of the HIRES only and
HIRES+SDSS data sets.  We note that, similarly to \cite{Viel:2005qj},
there is a preference for a non-zero $1/m_{\rm WDM}$ value which is at
present not statistically significant (less than $2\sigma$). The data
also prefers models with non-zero UV background fluctuations.

\begin{table}
\small
\caption{\small {Marginalized estimates (1$\sigma$ errors)}}
\label{tab2}
\begin{tabular}{lcc}
parameter & HIRES+SDSS & HIRES \\ 
\hline
\noalign{\smallskip}
  n               & $0.97 \pm 0.03 $  & $0.97 \pm 0.05 $  \\
  $\sigma_8$         & $0.96 \pm 0.07 $  & $1.0 \pm 0.2 $ \\
  $\Omega_{\rm m}$        & $0.25 \pm 0.03$& $0.28 \pm 0.09$ \\ 
  $\tau^A_{\rm eff} (z=3)$ & $0.35 \pm 0.01 $ & $0.33 \pm 0.03 $\\ 
  $\tau^S_{\rm eff} (z=3)$ & $3.17 \pm 0.07 $  & $3.02 \pm 0.37 $ \\
  $\gamma^A (z=3)$         &  $1.44 \pm 0.12 $ &  $1.54 \pm 0.33 $\\ 
  $\tau^A_{\rm eff} (z=5)$ & $1.53 \pm 0.09 $ & $1.54 \pm 0.19 $\\ 
  $\tau^S_{\rm eff} (z=5)$ & $4.77 \pm 0.44 $ & $4.92 \pm 0.5 $ \\
  $T_0  (z=3) (10^4)$ K   &  $2.23 \pm 0.30 $ &  $1.54 \pm 0.34 $\\
  $f_{\rm UV}$  & $0.65 \pm 0.25$ & $0.58 \pm 0.28$ \\  
  $1/m_{\rm WDM}$ (keV$^{-1}$)  & $0.09 \pm 0.07$   & $0.44 \pm 0.22$ \\  
\hline
\noalign{\smallskip}
\end{tabular}

\end{table}

{\bf{Discussion.}} We have used  a sample of high resolution \lya
forest spectra which is sensitive to the suppression in
the matter power spectrum at small scales caused by the free-streaming
of WDM particles.  We have modelled the observed flux power spectrum
by using high resolution hydro-dynamical simulations that incorporate
the relevant physical processes. We also improved previous
analyses by extending the parameter space, performing a Taylor
expansion order of the flux power spectrum in the cosmological and
astrophysical parameters to second instead of first order, and
including UV fluctuations that should be important at the high
redshifts considered here. We confirm that the observed \lya forest
flux power spectrum at small scales and high redshifts requires
significantly more power on small scales than provided by the models
that try to reproduce the cores of dwarf galaxies with a warm dark
matter particle by \cite{fornax}.  We improve previous lower limits on
the mass of warm dark particles of \cite{Seljak:2006qw} by a factor
two and those of \cite{Viel:2006kd} by a factor three. This further
decreases the rather small gap between the limits on the mass of DW
sterile neutrinos from \lya forest data and those on mass and mixing
angle from the diffuse X-ray background (e.g. \cite{asaka}). We note
that all the limits quoted here refers to the thermal production
mechanism for sterile neutrinos and could be potentially 10-20\%
weaker in case of non-thermality \cite{asaka}.

{\bf {Acknowledgments.}}  Simulations made at COSMOS funded by
PPARC, HEFCE, SGI, Intel and at  HPCF
(Cambridge, UK). GDB, MR and WLWS thank NSF for support under grants
AST 05-6845, AST-0206067 and AST-0606868. HIRES spectra obtained
at the Keck Obs. (supported by Keck Foundation). We thank
J.Lesgourgues for comments.

\end{document}